\documentclass[a4paper,fleqn,traditabstract]{aa}
\pdfoutput=1
\usepackage[pdftex]{graphicx}
\usepackage{epstopdf}
\usepackage{natbib,amssymb,amsmath,txfonts,astro,booktabs,textcomp}
\usepackage[version=3]{mhchem}
\usepackage[normalem]{ulem}
\DeclareGraphicsExtensions{.pdf,.png,.eps,.gif,.ps}
\begin{document}
\setlength{\mathindent}{0pt}
\def\phb#1{#1} 
\def\reac#1#2{\begin{equation}\label{r:r#1}\ce{#2}\end{equation}}
\def\reacnn#1{\[\ce{#1}\]}
\def\refreac#1{(\ref{r:r#1})}
\def\abS{\ensuremath{[\ce{S}]_{\rm tot}}}
\def\ab#1{\ensuremath{[\ce{#1}]}}
\def\fch2{\ensuremath{f_{\ce{CH2}}}}
\def\fhcn{\ensuremath{f_{\rm HCN}}}
\def\fhnc{\ensuremath{f_{\rm HNC}}}
\def\fcn{\ensuremath{f_{\rm CN}}}
\def\fifn{\ensuremath{\ce{^{15}N}}}
\def\fif{$^{15}$}
\def\fifnp{\ensuremath{\ce{^{15}N+}}}
\def\foun{\ensuremath{\ce{$^{14}$N}}}
\def\fifno{\ensuremath{\ce{$^{15}$NO}}}
\def\cfifn{\ensuremath{\ce{C$^{15}$N}}}
\def\nfifn{\ensuremath{\ce{N$^{15}$N}}}
\def\cfoun{\ensuremath{\ce{C$^{14}$N}}}
\def\hcfifn{\ensuremath{\ce{HC$^{15}$N}}}
\def\hfifnc{\ensuremath{\ce{H$^{15}$NC}}}
\def\conv{\ensuremath{{\cal N}\ul}}
\def\tauul{\ensuremath{\tau_{ul}}}
\def\ul{\ensuremath{{_{ul}}}}
\def\tul{\ensuremath{T_{ul}}}
\def\aul{\ensuremath{A_{ul}}}
\def\Ntot{\ensuremath{N_{\rm tot}}}
\def\qpart{\ensuremath{Q_{\rm exc}}}
\def\qtot{\ensuremath{Q_{\rm tot}}}
\def\qrot{\ensuremath{Q_{\rm rot}}}
\def\rr{\ensuremath{\thcn/\cfifn}}
\def\rrr{\ensuremath{\cfoun/\cfifn}}
\def\nratio {\ensuremath{{\foun/\fifn}}}
\def\nhhd   {\ensuremath{\rm NH_2D}}
\def\nnhp   {\ensuremath{\rm N_2H^+}}
\def\NH    {\ensuremath{\rm NH}}
\def\NHH    {\ensuremath{\rm NH_2}}
\def\nhhh   {\ensuremath{\rm NH_3}}
\def\fifnhh {\ensuremath{\rm ^{15}NH_2}}
\def\fifnhhh{\ensuremath{\rm ^{15}NH_3}}
\def\fifnnhp{\ensuremath{\rm ^{15}NNH^+}}
\def\nfifnhp{\ensuremath{\rm N^{15}NH^+}}
\def\perth{\ensuremath{\hbox{$\,^0\!/_{00}$}}}
\def\perth{\textperthousand}
\def\jnu{\ensuremath{J_\nu}}
\def\tpeak{\ensuremath{T_{\rm mb, 0}}}
\def\tpeak{\ensuremath{T_0}}
\def\aff{\ensuremath{A_{FF'}}}
\def\ratio#1{\ensuremath{{\cal R}(\ce{#1})}}
\def\ration#1{\ensuremath{{\cal R}_n(\ce{#1})}}
\def\refratio{{\ensuremath{{\cal R}_0}}}
\title{The CN/C$^{15}$N isotopic ratio towards dark clouds}
\author{%
P. Hily-Blant          \inst{1} \and
G. Pineau des For\^ets \inst{2} \and
A. Faure               \inst{1} \and
R. Le Gal              \inst{1} \and
M. Padovani            \inst{3}}

\institute{$^1$ Universit\'e Joseph Fourier and CNRS, Institut de
  Plan\'etologie et d'Astrophysique, Grenoble, France\\$^2$
  Universit\'e Paris-Orsay and CNRS, Institut d'Astrophysique
  Spatiale, Orsay, France\\$^3$ LRA/CNRS, Observatoire de Paris \& \'Ecole normale sup\'erieure, Paris, France}

\abstract{Understanding the origin of the composition of solar system cosmomaterials is a central question, not only in the cosmochemistry and astrochemistry fields, and requires various approaches to be combined. Measurements of isotopic ratios in cometary materials provide strong constraints on the content of the protosolar nebula. Their relation with the composition of the parental dark clouds is, however, still very elusive. In this paper, we bring new constraints based on the isotopic composition of nitrogen in dark clouds, with the aim of understanding the chemical processes that are responsible for the observed isotopic ratios. We have observed and detected the fundamental rotational transition of C$^{15}$N towards two starless dark clouds, L1544 and L1498. We were able to derive the column density ratio of C$^{15}$N over $^{13}$CN towards the same clouds, and obtain the CN/C$^{15}$N isotopic ratios, which were found to be $500\pm75$ for both L1544 and L1498. These values are therefore marginally consistent with the protosolar value of 441. Moreover, this ratio is larger than the isotopic ratio of nitrogen measured in HCN. In addition, we present model calculations of the chemical fractionation of nitrogen in dark clouds, which make it possible to understand how CN can be deprived of $^{15}$N and HCN can simultaneously be enriched in heavy nitrogen. The non-fractionation of \ce{N2H+}, however, remains an open issue and we propose some chemical way of alleviating the discrepancy between model predictions and the observed ratios.}


\maketitle
%

\section{Introduction}
\begin{table*}
  \centering
  \caption{Spectroscopic properties of the observed species and transitions.}
  \label{tab:spectro}
  \begin{tabular}{l c c c c c c c c}
    \toprule
    Molecule & $B^\dag$ & $\mu_0^\ddag$ & Transition$^{\dag\dag}$ & Frequency$^{\ddag\ddag}$  & $A_{ul}^\S$ & $g_u^\P$ \smallskip\\
    & MHz & Debye & & MHz & \pers & & \\
    \midrule
    \thcn  & 54353.130 & 1.45   &
    1\, 3/2\, 2\, 3 \ra 0\, 1/2\, 1\, 2 & 108780.201 & 1.05\tdix{-5} & 7 \smallskip\\
    \cfifn & 54958.510 & 1.45   & 
    1\, 3/2\, 2 \ra\ 0\, 1/2\, 1 & 110024.590 & 1.09\tdix{-5} & 5 \\
    & & &
    1\, 3/2\, 1 \ra\ 0\, 1/2\, 0 & 110023.540 & 7.16\tdix{-6} & 3 \\
    & & &
    1\, 3/2\, 1 \ra\ 0\, 1/2\, 1 & 110004.092 & 3.70\tdix{-6} & 3 \\
    & & &
    1\, 1/2\, 1 \ra\ 0\, 1/2\, 1 & 109689.610 & 7.10\tdix{-6} & 3 \\
    & & &
    1\, 1/2\, 1 \ra\ 0\, 1/2\, 0 & 109708.986 & 3.67\tdix{-6} & 3 \\
    & & &
    1\, 1/2\, 0 \ra\ 0\, 1/2\, 1 & 109733.657 & 1.08\tdix{-5} & 1 \\
    \bottomrule
  \end{tabular}
  \begin{list}{}{}
    \scriptsize
  \item $^\dag$ Rotational constant.
  \item $^\ddag$ Dipole moment.
  \item $^{\dag\dag}$ For \cfifn, the quantum numbers are $N J F$ with $J =
    N+S$ and $F=J + I$, where $S$ is the electronic spin and $I=1/2$ is
    the nuclear spin of $^{15}$N. For \thcn, we follow the CDMS
    convention (which differs from \cite{bogey1984}) where the quantum
    numbers are $N J F_1 F$ with $F_1=J + I_1$, $F=F_1 + I_2$ where
    $I_1=1/2$ and $I_2$ are the nuclear spin of $^{13}$C and $^{14}$N
    respectively.
  \item $^{\ddag\ddag}$ For resolved hyperfine structures, frequency of a given hyperfine component.
  \item $^\S$ Einstein coefficient for spontaneous decay.
  \item $^\P$ Total degeneracy of the upper level.
  \end{list}
\end{table*}

Understanding the origin of the elements in the solar system is a crucial issue at the crossroads of astrophysics, cosmochemistry, and astrobiology. In this respect, comets play a particular role because they are considered the most ancient relics of the formation of the solar system from the protosolar nebula, 4.567~Gyr ago. If it is certain that all cometary material has an interstellar origin, what is unclear is whether there is any chemical species present in \eg\ cometary ices that have remained unaltered since their incorporation into those ices. The central issue is to determine the origin of the elements that are detected in comets and interplanetary dust particles and to unveil their chemical link with the pristine interstellar matter. The detection of glycine in 81P/Wild~2 cometary samples by \cite{elsila2009} makes the question of the presence of glycine in the interstellar medium (ISM) still a crucial one. Direct analysis of interplanetary dust particles and meteorites provides clues to the composition of the solar system \citep[\eg ][]{bonal2010}, but due to their thermal and physical evolution since their formation, it is not always easy to trace the origins of the cosmomaterials back to the nascent protosolar system or even to the parental dark cloud. Isotopic composition may be a better tool in this respect. Isotopic ratios have proven extremely powerful in providing quantitative constraints on the composition of the protosolar disk from which the solar system formed \citep{chaussidon2007, jehin2009}. For carbon, the isotopic ratio \twc/\thc\ measured in comets shows a remarkable homogeneity, with a mean value of 91 that agrees with the solar value of 90, and is also very close to the value of 70 determined in the local ISM \citep{milam2005, asplund2009, manfroid2009}. For oxygen, measurements in comets lead to $^{16}$O/$^{18}$O ratios between 500 and 600, which again are in good agreement with the solar and local ISM value of 500 \citep{wilson1994}.

However, for nitrogen, which is the next most abundant element, measurements in various environments, such as the Sun, the solar system, and the ISM, draw a much more contrasted picture. The Genesis mission has recently allowed \cite{marty2011} to measure the nitrogen isotopic ratio in the solar wind, yielding $\nratio = 441\pm5$ for the protosolar nebula. This value is significantly larger than the terrestrial value of 272. A review of the \nratio\ measurements in primitive matter of the solar system \citep{mumma2011} confirms the idea of a strongly variable \nratio\ ratio in the solar system. Observations of the \nratio\ ratio in protoplanetary disks may also be directly compared to the value from the Genesis mission. Unfortunately, measurements of isotopic ratios in the most evolved gas-poor disks like TW~Hya \citep{kastner2010} are extremely difficult, except for carbon and oxygen-bearing molecular species \cite[\eg\ CO][]{dutrey1997chem, dutrey2007} whose large abundances make their major isotopologues detectable. This is in contrast with cold and dense starless cores, where N-bearing molecules (HCN, HNC, CN, \nhhh, \nnhp, NO) are routinely observed \citep{gerin1992, crapsi2005, crapsi2007, hilyblant2010n, padovani2011}, together with their \fifn\ isotopologues \citep{ikeda2002, gerin2009, lis2010}. The recent observation of \nfifnhp\ towards the starless core L~1544 by \cite{bizzocchi2010} leads to an isotopic ratio \nnhp/\nfifnhp $=446 \pm 71$, a value equal to that of the protosolar nebula. Similar values were obtained by \cite{gerin2009} in other cold dense cores using \nhhd, though with much larger uncertainties. These values contrast with the HCN measurements by \cite{ikeda2002} and \cite{lucas1998}, who find ratios of 150 and 240 in dark and diffuse molecular gas respectively. More recently, \phb{\cite{adande2012} found similar values based on HCN and CN observations towards warm ($T>25$~K) molecular clouds, with ratios from 150 to 350.} Even more recently, \cite{hilyblant2013ica} (hereafter HB13) report HCN isotopic ratios between 140 and 360, where the range of values is shown to reflect genuine spatial variations of the ratio. HB13 also suggest that the large variations of the \nratio\ ratio in cosmomaterials, in dark clouds, and in the more diffuse ISM may indeed be rooted in the interstellar nitrogen chemistry. The underlying idea is that there are two reservoirs of interstellar nitrogen characterized by different isotopic ratios: nitrile carriers (such as HCN) are systematically enriched in \fifn\ with respect to the Sun's value, whereas hydrides (such as ammonia) show values encompassing that measured in the Sun. The differential fractionation of these two reservoirs may originate in the interstellar chemistry of nitrogen. This is because nitriles essentially derive from atomic nitrogen, whereas hydrides derive from molecular nitrogen. 

Because atomic and molecular nitrogen, which are presumably the main reservoirs of nitrogen, are not directly observable in the dense ISM, the determination of the bulk \nratio\ is indirect and relies on the observation of minor N-bearing molecules. To which extent the abundance ratio of two isotopologues of a given molecule is representative of the elemental isotopic ratio depends on so-called fractionation processes: the stronger these processes, the larger the departure of the molecular abundance ratio from the elemental ratio. Fractionation processes are of two types: chemical and selective photodissociation effects. The latter effect can safely be ignored in dark clouds, which are dense and shielded regions. Chemical effects are potentially important in cold gas \citep{watson1976, langer1984}. Regarding nitrogen, \cite{terzieva2000} (hereafter TH00) investigated chemical fractionation in dense and cold clouds (\nh=2\tdix{4}\,\ccc, $T=10$~K), and found essentially no fractionation of any nitrogen-bearing molecule. This was further investigated by \cite{charnley2002}, who showed, in contrast, that in situations where OH is strongly depleted from the gas-phase, there is more atomic nitrogen available for fractionation. Indeed, large fractionation levels are predicted, \eg\ for \ce{NH3}, at early times in their calculations. However, these models also predict strong fractionation of \ce{N2H+}, which is not observed. Therefore, none of the two available models of nitrogen fractionation in dark clouds are able to reproduce the observational constraints.

The CN radical plays a peculiar role in the nitrogen chemistry because it is a mediator of the atomic-to-molecular conversion (HB13). CN is also coupled with the dominant nitriles such as HCN and HNC, which derive from atomic nitrogen. Thus, the nitrogen fractionation level of CN potentially provides interesting constraints on the fractionation pathways. In this paper, we report the detection of \cfifn\ towards two starless cores, L1544 and L1498. Put in conjunction with the \ce{N2H+} and HCN fractionation measurements available towards L1544, it is expected that these new observations will strongly constrain models.

The paper is structured as follows. In Section 2, we present the observations and column density derivation. In Section 3, we present our fractionation model, to which we compare the \nratio\ measured in CN and other molecules. The discussion is the subject of Section 4.

\section{Observations}

The $N=1-0$ rotational spectrum of \cfifn\ is split into fine and hyperfine structures, which arise from the electron spin and the $I=1/2$ nuclear spin of the \fifn\ atom. There are two groups of lines due to the fine structure, which are distant by 300~MHz (see Table~\ref{tab:spectro}). The strongest hyperfine component is the $(N,J,F)=(1,\frac{3}{2},2)\ra (0,\frac{1}{2},1)$ transition at 110024.590~MHz, with a relative intensity of 0.417 \citep{saleck1994}. The next strongest transitions have relative intensities of 17\%, with frequencies of 109689.610 and 110023.540~MHz. All six hyperfine components were observed simultanously towards L1498 and L1544 at the IRAM 30m telescope in May 2012 ($\approx 16$~hrs per source) under good weather conditions. The VESPA autocorrelator was used as a spectrometer to provide us with high-resolution spectra, namely, 20~kHz or 0.054~\kms\ at 110~GHz. The typical receiver temperature achieved with the EMIR detectors \citep{carter2012} was $\approx 40$~K, resulting in system temperatures $\approx 120$~K. Amplitude calibration was performed using the three-phase IRAM method typically every 10~min. Pointing and focus sessions using strong continuum sources were repeated every one and two hours respectively. The wobbler-switching mode (140\arcsec\ throw) was used to provide us with high-quality baselines, such that residual bandpass calibration effects could be handled with first-order polynomial fitting. The \hcfifn(1-0) data reported in this paper are taken from \cite{hilyblant2010n}. The \thcn(1-0) rotational spectrum also presents fine and hyperfine structures, which are described in \cite{bogey1984}. For the present analysis, we use only the brightest hyperfine component, at 108780.201 \citep{hilyblant2008cn}. The spectroscopic properties of both species and observed transitions are summarized in Table~\ref{tab:spectro}. Towards L1544, the \thcn\ spectrum is taken from \cite{hilyblant2010n}, and towards L1498, it is taken from unpublished spectra of \cite{padovani2011}. Data reduction was done using the CLASS software \citep{hilyblant_class}. All plots were done in the GILDAS environment provided by IRAM \citep{pety_gildas}. The J2000 coordinates for L1498 and L1544 are $\alpha, \delta=$04:10:51.50, 25:09:58.0 and 05:04:16.90, 25:10:47.0 respectively.

\section{Results}

Among the three brightest \cfifn\ lines, only the strongest, with a relative intensity of 42\%, is clearly detected towards both sources. The line properties, which are derived from Gaussian fits, are summarized in Table~\ref{tab:results}. We also give the properties of the \thcn(1-0) line obtained in a similar fashion. We note a 0.1\kms\ discrepancy between the velocities of the \thcn\ and \cfifn\ lines towards L1498, which could not be ascribed to any obvious technical issue. \phb{We note, however, that the uncertainty on the rest frequency reported in the Cologne Database for Molecular Spectroscopy \citep{muller2005} is 0.1~MHz, or 0.25\kms, which could account for the observed shift.} New observation of the \thcn(1-0) towards L1498 should be performed to check this discrepancy. We are, however, most confident that the detected lines are \thcn\ and \cfifn, since we searched for other lines at those frequencies in public databases and could not find any likely candidate. There is a marginal detection of the 110023.540~MHz line towards L1544, which has a peak intensity of 20~mK. This intensity well matches what would be expected from the relative intensities (17\% that of the strongest line), which suggests that the lines are optically thin. This secondary component is not detected towards L1498, though there is some emission \phb{at the expected velocity (within the 0.1~MHz uncertainty)}, with an upper limit that is consistent with the relative intensities of these two components. The analysis below is based solely on the strongest \cfifn\ line at 110024.590~MHz.

Line intensities towards L1544 are stronger than towards L1498 and are also broader by a factor $\approx 2$. This is likely due to the L1544 lines presenting two closeby velocity components \citep{hilyblant2010n}, which are not resolved in the \cfifn\ spectrum. These two velocity components may be related to the collapse signature recently evidenced by the inverse P-Cygni profile of a water line \citep{caselli2012}.

To derive column densities from the emission lines, we performed
radiative transfer calculations with the \texttt{RADEX} code
\citep{vandertak2007}, using the large velocity gradient (LVG)
approximation for an expanding sphere. The hyperfine rate coefficients
for C$^{15}$N and $^{13}$CN were derived from the rate coefficients
computed by \cite{kalugina2012} for CN+H$_2$ using the
infinite-order-sudden approximation (see Appendix~C). We assumed that
the source has a constant density and temperature and that it
uniformly fills the telescope beam. The line width was held fixed at
0.2~\kms\ for L1498 and at 0.4\kms\ for L1544. The intensity of the
observed \thcn(1-0) transition was computed for varying \thcn\ column
density ($N(\thcn)$), \hh\ density (\nhh), and kinetic temperatures
($T$). The kinetic temperatures considered span the range 7 to 12~K by
steps of 1~K. Figure~\ref{fig:lvg} shows the resulting line intensity
for a kinetic temperature of 10~K. A continuum range of solutions
matches the observation constraints summarized in
Table~\ref{tab:results}, with densities $\nhh=\dix{4}$ to a few
\dix{5}\ccc\ and associated \thcn\ column densities from 6\tdix{13} to
2.5\tdix{12}\cc. Refined calculations were then performed across this
range for both \thcn\ and \cfifn. For \cfifn, a similar ensemble of
solutions was found. Even though the density and column density are
degenerate, the \thcn\ and \cfifn\ solutions are characterized by a
constant \thcn/\cfifn\ column density ratio across the full range of
solutions. Finally, refined calculations were done in which the
\cfifn\ column density was such that $6 \le N(\thcn)/N(\cfifn) \le
9$. The \thcn/\cfifn\ column density ratios are 7.5 and 7 for L1544
and L1498 respectively. Although the density and column density are
degenerate, equal excitation temperatures (within 0.1~K) of $3-5$~K
are found for the \thcn\ and \cfifn, across the full range of
solutions. The \thcn\ line is found to be only marginally thin, with
an opacity close to unity, whilst the \cfifn\ line can safely be
assumed to be thin. It is of interest to note that LTE calculations
using excitation temperatures in the range 3--4~K provide abundance
ratios which agree reasonably well with the above LVG calculations.

On chemical grounds, both species follow similar formation and
destruction routes and are thus expected to coexist spatially hence
the above column density ratios are representative of their abundance
ratios. If we now assume that \twc N/\thcn=\twc/\thc\ and adopt
for the latter the value of 68 from \cite{milam2005}, we derive an
isotopic ratio $\ce{CN}/\cfifn=476\pm 70$ in L1498 and
$\ce{CN}/\cfifn=510\pm 70$ in L1544, where conservative uncertainties
are applied (see Table~\ref{tab:results}). Altogehter, these values
give $\ce{CN}/\cfifn=500\pm 75$.

The \thcn\ column density may also be compared with the H\thcn\ column
density, which is 1.8\tdix{12}\cc\ towards L1498 \citep{padovani2011}
and 4.6\tdix{12}\cc\ towards L1544 (HB13). The \thcn:H\thcn\ ratios
are thus 0.6 and 1.7 for L1498 and L1544 respectively, which are in
harmony with ratios found by \cite{hilyblant2010n} towards several
starless cores.

\begin{table*}[t]
  \centering
  \caption{Line properties towards L1544 and L1498 and associated column density ratios from LVG calculations.}
  \begin{tabular}{l cccc cccc cc}
    \toprule
    Source
    & \multicolumn{4}{c}{\thcn$^\S$}
    & \multicolumn{4}{c}{\cfifn$^\S$}
    & \thcn/\cfifn$^\dag$
    & \cfoun/\cfifn$^\ddag$ \\
    \cmidrule(rl){2-5}
    \cmidrule(rl){6-9}
    &  $W$ & $\tpeak$ & $v_0$ & FWHM
    &  $W$ & $\tpeak$ & $v_0$ & FWHM\\
    & m\kkms & mK & \kms & \kms
    & m\kkms & mK & \kms & \kms \\
    \midrule
    L1498    & 28(2) & 114(14)& 7.84(0.01) & 0.23(0.02)
             &  6(1) & 34(7)  & 7.73(0.02) & 0.17(0.03)
             & 7.0$\pm$1.0 & 476$\pm$70 \\
    L1544    & 73(4) & 166(15)& 7.17(0.02) & 0.44(0.03)
             & 20(2) & 47(10) & 7.18(0.03) & 0.40(0.05)
             & 7.5$\pm$1.0 & 510$\pm$70 \\
    \bottomrule
  \end{tabular}
  \begin{list}{}{}
  \item $\S$ From Gaussian fits. Statistical
    uncertainties at the 1$\sigma$ are in parenthesis for a velocity
    resolution of 0.05\kms\ at 110~GHz.
  \item $\dag$ Adopting FWHM of 0.20 an
    0.40\kms\ for L1498 and L1544 respectively. Conservative
    uncertainties based on $\chi^2$ minimization.
  \item $\ddag$ Adopting \twc/\thc=68 \citep{milam2005}.
  \end{list}
  \label{tab:results}
\end{table*}

\section{Comparison to model predictions}

\subsection{Fractionation of nitriles and hydrides}

The CN/\cfifn\ abundance ratios inferred in the two starless cores are thus \phb{only slightly larger than the protosolar value of \nratio=441, but significantly larger the HCN/HC\fifn\ ratio derived by HB13 towards L183 and L1544, which are smaller than the protosolar ratio}. In contrast, the CN/\cfifn\ ratio is similar to that measured in hydrides (\ce{NH3}, \ce{N2H+}) towards dark clouds by \cite{bizzocchi2010}, \cite{gerin2009}, and \cite{lis2010}. At first glance, this is at odds with the suggestion that hydrides and nitriles present different fractionation degrees (HB13), because one would have expected CN to behave more like HCN than like \ce{NH3}. Yet, it must be recognized that CN plays a particular role in the chemistry of nitrogen: though it is a nitrile, it reacts directly with N to form \ce{N2} from which hydrides derive. Hence, the non-fractionation of CN may not be in contradiction with the chemical picture proposed in by HB13. 

\subsection{Model calculations}

We thus wish to understand how the CN radical gets deprived in \fifn\ and at the same time HCN is enriched in \fifn. To this aim, we implemented the fractionation reactions of TH00 into a dark cloud chemical network based on \cite{flower2003}, in which the nitrogen chemistry has been fully revised. The most important changes regard the dissociative recombinations of ions such as \ce{NH2+}, \ce{NH3+}, and \ce{NH4+}, which are critical for the production of hydrides. Following \cite{dislaire2012}, the dissociative recombination of \ce{N2H+} possesses a channel producing NH with a branching ratio of $\approx 5$\% \phb{\citep{vigren2012}}. \phb{The gas phase ortho-to-para conversion of \ce{H2} through proton exchange reactions has also been fully updated \citep[][and references therein]{rist2013} since the rate of the key reaction \ce{N+ + H2} depends critically on the ortho-to-para ratio of \ce{H2} \citep{dislaire2012}}. Other important reactions have been updated, based on the recent laboratory work of \cite{daranlot2012}. In a first approach, the updated chemical network of Le Gal was simplified by ignoring the ortho- and para- forms of \hh, \ce{H3+}, and other nitrogen-bearing molecules. This means that the rate of the reaction \ce{N+ + H2} was held fixed at a value corresponding to an ortho:para ratio of \hh, of \dix{-3}, in a similar fashion as in \cite{dislaire2012}.

The chemical fractionation reactions considered by TH00 are ion-neutral reactions, because they are the fastest and are hence expected to be the most efficient. These reactions consist in forward and backward reactions,
\reacnn{A+ + B <=>[k_f][k_r] C+ + D + $\Delta E,$}
where $\Delta E$ is the zero-point energy difference between the products and reactants, such that $\Delta E > 0$ if the reaction is exothermic. The rates of the forward and backward reactions are related through
\[
k_f / k_r = f(B,m) \exp[\Delta E / kT],
\]
where $f(B,m)$ is a factor that depends on the masses and partition functions of the reactants and products. The values of $\Delta E$ and $f(B,m)$ are taken from the work of TH00 and are listed in Table~\ref{tab:reactions}. At temperatures smaller than $\Delta E$, the forward reaction proceeds much faster than the backward reaction, thus shifting the steady-state abundances towards the reactants. In implementing the fractionation reactions of TH00 in our network, we adopted a single Langevin rate of \dix{-9}\,\cccs\ for all forward reactions. The remaining reactions involving nitrogen atoms in our network have been duplicated to include their \fifn\ homologue. The total number of species and reactions are 144 and 870 respectively. We do not consider freeze-out of any species in these models. We follow the time evolution of the chemical state of a parcel of gas of density (\nh=\dix{4}\ccc) and kinetic temperature of 10~K, which was exposed to a cosmic-ray ionisation rate $\zeta=1.3\tdix{-17}\pers$ and shielded from the ambient UV field by 20~mag of visual extinction, such that external UV photons do not play any significant role. However, secondary photons, which result from the fluorescence of \ce{H2} molecules excited by collisions with the electrons produced by cosmic-rays are explicitly taken into account and are responsible for a significant fraction of the ionisation. The initial abundances are given in Table~\ref{tab:abinit}, where an elemental isotopic ratio of nitrogen $\refratio =  440$ was adopted \citep{marty2011}.

\begin{figure*}
  \centering
  \includegraphics[width=0.48\hsize]{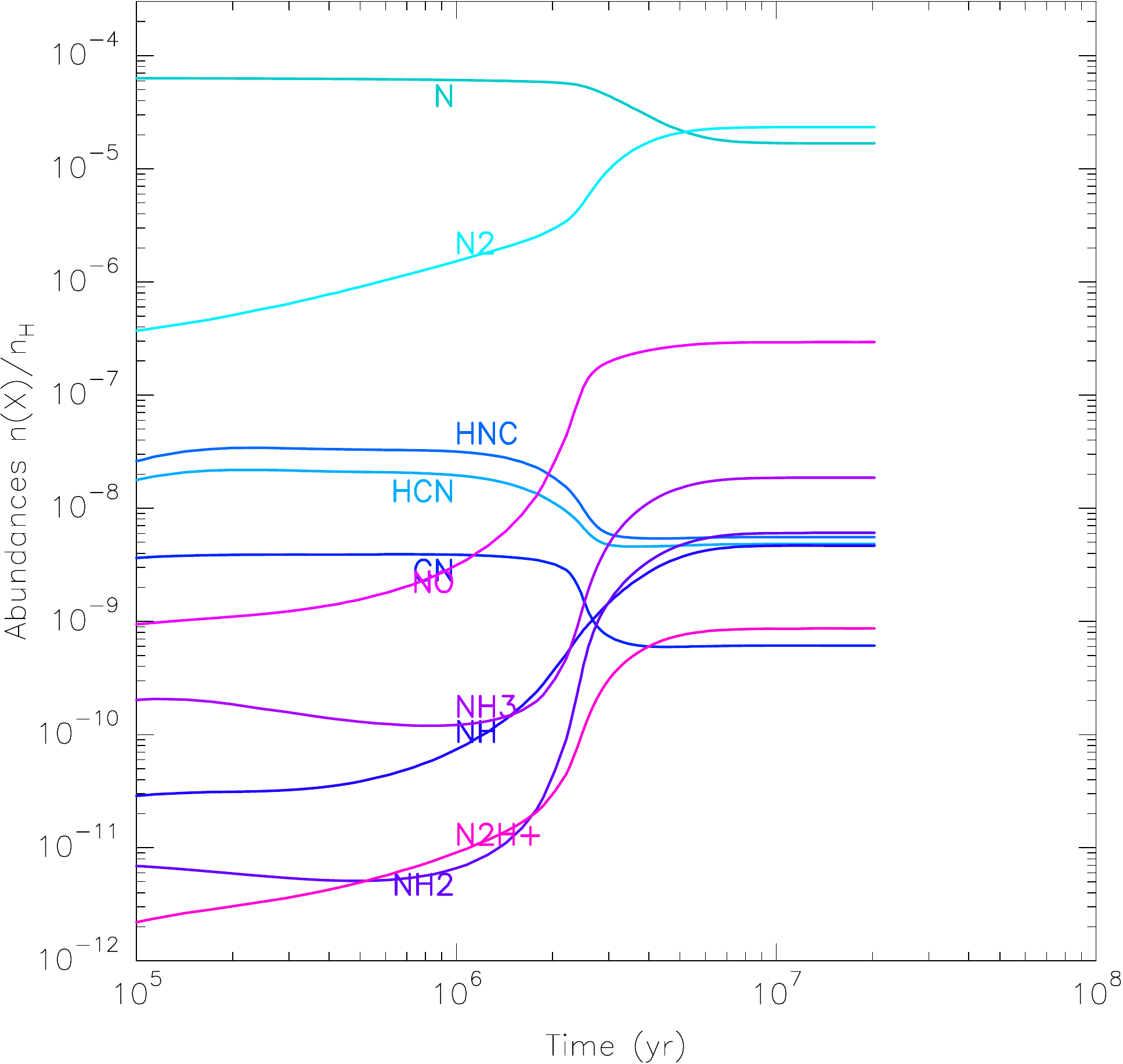}\hfill%
  \includegraphics[width=0.48\hsize]{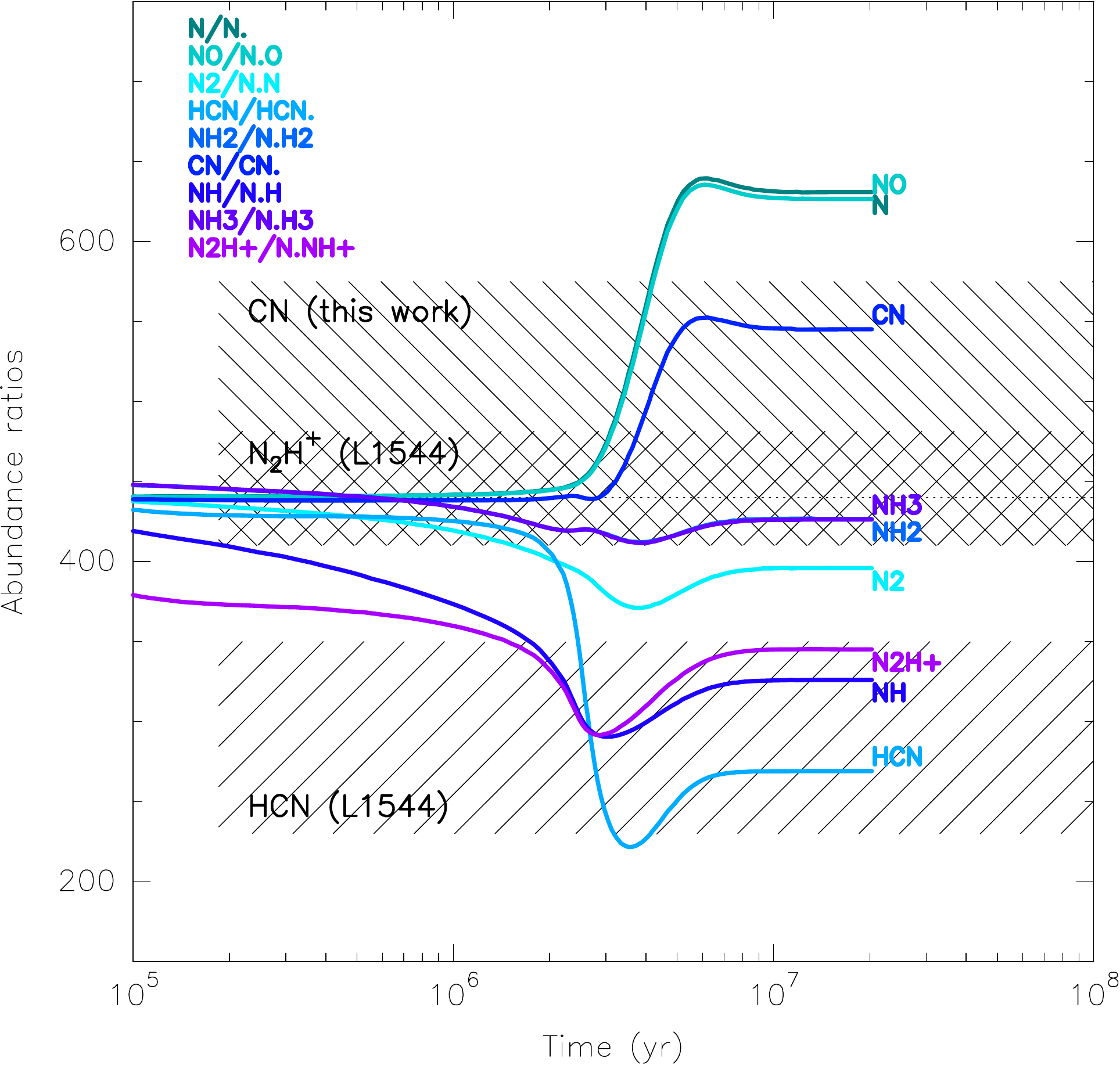}
  \caption{Fractional abundances (left) and abundance ratios (right)
    of selected species. The observed isotopic ratios towards L1544
    for HCN and CN, as well as towards L1498 for CN, are shown as
    hatched areas (right panel). The \fifn\ atom is noted N.,
    such that CN. matches \cfifn, etc. The dotted line shows the
    adopted reference elemental isotopic ratio, \foun/\fifn=440
    \citep{marty2011}.}
  \label{fig:model}
\end{figure*}

\subsection{Chemical considerations}

The resulting abundances and abundance ratios of selected species are shown in Fig.~\ref{fig:model}. \phb{As is evident, all species reach a steady state after typically 5~Myrs, similar to what was found by \cite{flower2006b}. This timescale is comparable to the lifetime of molecular clouds \citep{tassis2004} but is larger than the free-fall timescale and slightly larger than the ambipolar diffusion one \citep[\eg][]{walmsley2004}. Those dynamical timescales are mostly relevant for the inner and dense parts of dark clouds and are not directly applicable to the rather low density material modelled here. It is possible, but not certain, that the nitrogen-bearing species have reached steady-state. In any case, it must be recognized that the chemical timescale depends strongly on the initial abundances adopted.}

In our calculations, the steady state is characterized by an N:\ce{N2} abundance ratio close to unity, suggesting near equipartition between these two reservoir species. This is in sharp contrast to the calculations of \cite{hilyblant2010nh}, who found \ce{N2} to be 30--40 times more abundant than N. Unfortunately, the use of two different chemical networks, namely OSU 2009 in the latter case, impedes any direct comparison but undoubtedly calls for further investigation. Still, our predicted ratio is similar to that determined by \cite{flower2006b} with an older version of our chemical network. Another characteristic is the HNC:HCN$\approx 1$ ratio. We also see that \ce{NH3} is the most abundant of the three hydrides, with an abundance of 1.2\tdix{-8}, when NH and \ce{NH2} have abundances of 5.1\tdix{-9} and 4.3\tdix{-9} respectively. The predicted abundance of ammonia is close to its value of 1.5-5.0\tdix{-8} in the cold envelope of IRAS16293-2422. The latter estimate is a revision of the value published in \cite{hilyblant2010nh}. It is based on non-LTE radiative transfer calculations using the publicly available \texttt{RATRAN} and \texttt{LIME} codes, and taking into account all the observed ammonia lines, including the 572~GHz fundamental line \citep{ratran, lime}. These calculations will be detailed elsewhere. There is also a good agreement for the abundances of NH and \ce{NH2}. The predicted CN:HCN abundance ratio of 0.2 is smaller than the observed value, a discrepancy already noted by \cite{hilyblant2010n}.


We now examine the predicted fractionation degree, that is, the enrichment or depletion in \fifn\ with respect to the elemental ratio, of several nitrogen-bearing species, which is our main concern. The right panel of Fig.~\ref{fig:model} shows the abundance ratios of nitrogen-bearing isotopologues as a function of time. Probably the main result is that our model does predict large variations of the isotopic ratio in several species, contrary to the calculations of TH00, who found very small fractionation for all species. Our model predicts that the isotopic ratio of \ce{N}, which we note \ratio{N}, \phb{is $\approx 630$, and that $\ratio{N2} \approx 400$} (see Table~\ref{tab:model}). Since these species are also the reservoirs of nitrogen, \phb{this implies that most of the nitrogen is fractionated, either enriched or deprived in \fifn\, and any enrichment/depletion of N in \fifn\ implies depletion/enrichment of \ce{N2}}. Another striking feature is that until a few \dix{6}~Myrs, isotopic ratios remain close to the initial value of 440. Beyond this time, three types of behaviour may be emphasized: species that are enriched in \fifn, those that are deprived in \fifn, and those that are not fractionated. The latter group includes \ce{NH2} and \ce{NH3}. The positively fractionated molecules are HCN, HNC, and NH, for which the steady-state isotopic ratio is $\approx 330$, and the depleted ones are N, NO, and CN. (see Table~\ref{tab:model}). The latter exhibits an isotopic \phb{ratio of $\approx 550$}.

The depletion of atomic nitrogen in \fifn\ is due to the fractionation reactions
\reac{fracNa}{^{15}N + N2H+ -> ^{15}NNH+ + N}
and
\reac{fracNb}{^{15}N + N2H+ -> N^{15}NH+ + N}
that are strongly favoured at low temperature with respect to the reverse process, due to a large $\Delta E = 36$~K (see Table~\ref{tab:reactions}). The ratios predicted for N and NO are very close, and that of CN, although smaller, follows the same trend. The former result is easily explained by recognizing that NO and \fifn O are produced and destroyed by the same two reactions:
\reac{formNO}{N + OH ->[f_1] NO + H}
\reac{destNO}{NO + N ->[d_1] N2 + O}
For \fifn O, the main formation channel is through reaction with \fifn, whilst the destruction channel is through reaction with N. Hence the steady-state abundances of NO and \fifn O are given respectively by
\[
{[\ce{NO}]} = f_1/d_1 \times [\ce{OH}]
\]
and
\[
{[\fifno]} = f_1/d_1 \times [\fifn][\ce{OH}] / [\ce{N}],
\]
leading to an isotopic ratio NO/\fifn O = N/\fifn. The situation is similar for CN, but as already noted, the isotopic ratio of CN is lower than that of N (and NO). This is because \ce{C2N+} and \ce{H2CN+} are efficiently fractionated by the ion-neutral reactions of TH00 (see Table~\ref{tab:reactions}), such that their isotopic ratios become smaller than \refratio. The dissociative recombination of these ions, which are normally negligible in producing CN, contributes to about 10\% to the formation of \cfifn. As a result, \cfifn\ does not strictly follow the same behaviour as N and NO. Because HCN and HNC are primarily formed through the dissociative recombination of \ce{H2CN+}, their isotopic ratios are similar to that of \ce{H2CN+}.

We now turn our attention to nitrogen hydrides. To understand the different behaviour of NH on the one hand and \ce{NH2} and \ce{NH3} on the other hand, we first need to understand what determines the relative abundances of \fifnp\ and \ce{N+}. The \ce{N+} ion is mostly produced by the reaction of \ce{N2} with \ce{He+}. This is also true for \fifnp, but the fractionation reaction
\reacnn{\fifnp + N2 -> N+ + N^{15}N}
presents an exothermicity of 28~K. The forward process leading to \ce{N^{15}N} is therefore enhanced at low temperature with respect to the reverse process, and this is why $\ratio{N2} <  \ratio{N+}$. As \ce{N2H+} is a daughter molecule of \ce{N2}, this also explains how $\ratio{N2H+} < \refratio$, in addition to reactions \refreac{fracNa}-\refreac{fracNb}. In turn, the enrichment of \ce{N2H+} in \fifn\ propagates into NH because, as \cite{dislaire2012} have shown, the dissociative recombination of \ce{N2H+} dominates the formation of NH over the \ce{NH2+ + e-} channel. Therefore, we find that $\ratio{NH} < \refratio$. To explain how \ce{NH2} and \ce{NH3} behave differently, we recall that these two molecules result from the dissociative recombination of \ce{NH4+}, which itself results from successive hydrogen abstraction reactions initiated by \ce{N+ + H2}. The fractionation of \ce{NH2} and \ce{NH3} therefore reflects that of \ce{N+}. Hence, \ratio{NH} reflects that of \ce{N2}, which is smaller than \refratio, whereas \ratio{NH2} and \ratio{NH3} reflect that of \ce{N+}, which is larger than \refratio.

However, as is evident from Fig.~\ref{fig:model} and Table~\ref{tab:model}, \ratio{NH2} and \ratio{NH3} are not strictly equal to \ratio{N+}, as we would expect from the above reasoning. Indeed, the fractionation levels of \ce{N+}, \ce{NH+}, \ce{NH2+}, \ce{NH3+}, and \ce{NH4+}, are 480, 470, 450, 434, and 429 respectively. This shows that there is some enrichment in \fifn\ at each step along the reaction chain which begins with \ce{N+ + H2} and terminates with \ce{NH3+ + H2}. The reason is that, because $\ratio{NH}>\refratio$, the reaction \ce{H+ + ^{15}NH} is faster than the similar reaction \ce{H+ + NH} and thus becomes non-negligible in the formation of \ce{^{15}NH+}. As a consequence, $\ratio{NH+} < \ratio{N+}$. In a similar fashion, the two reactions \ce{H3+ + ^{15}NH} and \ce{HCO+ + ^{15}NH} contribute to the formation of \ce{^{15}NH2+}, whereas their \foun-homologues do not. This explains how the positive fractionation of NH propagates in the other two hydrides, though at a small level. Now, the next ions in the sequence, namely \ce{^{15}NH3+} and \ce{^{15}NH4+}, are further enriched in \fifn\ with respect to \ce{NH2+} because \ce{^{15}NH3} starts to contribute to their synthesis by reacting with \ce{H+}, \ce{H3+}, and \ce{C+}. In contrast, the \foun-homologues of these reactions are negligible in forming \ce{NH3}. As we see, minute variations of the isotopic ratio of a given species propagate into the chemical network because the increase in the relative abundance of the \fifn-isotopologue enhances reactions that are usually negligible, thus opening new formation paths.

Finally, the above chemical considerations may be summarized as follows. Atomic nitrogen is depleted in \fifn\ with respect to \ce{N2} because of the fractionation reactions \refreac{fracNa}-\refreac{fracNb}. Other reactions decrease \ratio{H2CN+} below \refratio, with the consequence that both \ratio{HCN} and \ratio{HNC} are lower than \refratio. Species that derive primarily from atomic nitrogen, like NO and CN (see \eg\ HB13), have isotopic ratios close to that of N. Yet, for CN, $\ratio{C2N+} < \refratio$ drives \ratio{CN} to values lower than \ratio{N} and \ratio{NO}. Among hydrides, the positive fractionation of NH derives from that of \ce{N2H+}, which is inherited from that of \ce{N2}. In contrast, \ce{NH2} and \ce{NH3} are neither enriched nor depleted in \fifn\ because they derive principally from \ce{N+}, which is only very slightly fractionated.

Since the rate of the key reaction \ce{N+ + H2} depends on the ortho:para ratio of \hh, we performed a series of calculations with ortho:para ratios of \hh\ ranging from \dix{-6} to \dix{-3}. The predicted isotopic ratios do not change by more than 10\%, except those of \ce{NH2} and \ce{NH3} which follow the same trend, decreasing from 427 to 305 when o:p is \dix{-3} and \dix{-6} respectively. In the process, the ratio of NH decreases only from 326 to 317. The reason is that as o:p of \hh\ decreases, the rate of the reaction \ce{N+ +H2} drops and the reaction \ce{NH + H+} starts to dominate the formation \ce{NH+} and \ce{^{15}NH+}. This explains how the ratios of \ce{NH2} and \ce{NH3} become closer to that of NH. Apart from this moderate change affecting \ce{NH2} and \ce{NH3}, we conclude that the above results do not depend on the assumed value of o:p of \dix{-3}.

\subsection{Comparisons to the observations}

The present model predictions are now compared to the CN and HCN isotopic ratios observed in L1544 and L1498 (right panel of Fig.~\ref{fig:model}). The isotopic ratio of HCN are $\ratio{HCN}=230-350$ and 190--280 towards the central positions of L1544 towards L183 respectively (HB13). This value is not available in the case of L1498. For CN, we take the values determined in Section 3. As is evident, the model predictions agree well with all the observational constraints. In particular, the isotopic ratios of CN and HCN estimated in L1544 are well reproduced by our model calculations.

More problematic is the case of \ce{N2H+}. The constraints on \ce{N2H+}/\ce{N^{15}NH+} from \cite{bizzocchi2010}, who found 446$\pm$71, are plotted in Fig.~\ref{fig:model}. Our model predicts a significant enrichment for \ce{N^{15}NH+}, with $\ratio{N2H+}=280$. The fractionation of \ce{N^{15}NH+} is associated with the reaction of \ce{N2H+} with \fifn. From Table~\ref{tab:model}, we note that the fractionation of \ce{N^{15}NH+} is larger than that of \ce{^{15}NNH+}, a result of the different zero-point energy differences of the two reactions
\reacnn{N2 + N^{15}NH+ -> ^{15}NN + N2H+}
and
\reacnn{N2 + ^{15}NNH+ -> ^{15}NN + N2H+ ,}
whose exothermicities are respectively $\Delta E=10.7$ and 2.25~K. These reactions may proceed either by breaking an N-H bond or an N-N bond. In the latter case, the reaction likely requires a strong activation energy. If we artificially switch off these two reactions, we note that the fractionations of \ce{^{15}NNH+} and \ce{N^{15}NH+} become equal, as expected, with a isotopic ratio of 340. Yet, this value is only slightly larger than the value in Table~\ref{tab:model} and is still outside the permitted value from \cite{bizzocchi2010} by a small factor. The same argument may apply even more likely to the fractionation reactions \ce{N + N2H+}, which proceeds through N-N bond breaking. Suppressing the latter reactions has a much stronger effect on the fractionation picture because these reactions are responsible for the fractionation of atomic nitrogen (see the discussion above). In particular, NH becomes less fractionated and CN and \ce{N2H+} have similar isotopic ratios $\approx 400$, whereas HCN fractionation remains unaffected. This further illustrates the dichotomy suggested by HB13. Observations of \ce{^{15}NNH+} would enable the first suggestion to be tested.

{The models presented here have a C/O abundance ratio of 0.50, which falls in between the low--metal-abundance value of 0.41 from \cite{wakelam2008} and the value of 0.67 from \cite{flower2003}. We also performed model calculations with C/O=0.40, for which all the above reasoning applies equally, though predicting a smaller \ratio{CN} whilst maintaining a fractionation of HCN close to the case C/O=0.50. We thus found a better agreement with the calculations having C/O=0.50. Exploring the effects of the initial abundances is an important area of investigation. However, it is beyond the scope of the present work and is deferred to a subsequent paper.

\begin{table}[t]
  \caption{Steady-state abundances of selected species and their isotopologues
    (see also Fig.~\ref{fig:model}).}
  \label{tab:model}
  \begin{center}
    \begin{tabular}{l r l r c c c}
      \toprule
      Species & [X]$^\S$ & Species & [X]$^\S$ & ${\cal R}^\dag$ & ${\cal R}_n^\ddag$ & $\delta_{15}^\sharp$\\
      \midrule
%
     \ce{N}&   1.7(-05)&     \ce{^{15}N}&   2.7(-08)&  631&  1.4& -569\\
    \ce{N2}&   2.3(-05)&    \ce{^{15}NN}&   1.2(-07)&  396&  0.9& -313\\
    \ce{NO}&   2.9(-07)&    \ce{^{15}NO}&   4.7(-10)&  627&  1.4& -566\\
    \ce{CN}&   6.1(-10)&    \ce{C^{15}N}&   1.1(-12)&  545&  1.2& -501\\
   \ce{HCN}&   4.8(-09)&   \ce{HC^{15}N}&   1.8(-11)&  269&  0.6&   11\\
    \ce{NH}&   4.7(-09)&    \ce{^{15}NH}&   1.4(-11)&  326&  0.7& -166\\
   \ce{NH2}&   6.1(-09)&   \ce{^{15}NH2}&   1.4(-11)&  427&  1.0& -363\\
   \ce{NH3}&   1.9(-08)&   \ce{^{15}NH3}&   4.4(-11)&  426&  1.0& -362\\
  \ce{N2H+}&   8.6(-10)&  \ce{^{15}NNH+}&   2.5(-12)&  345&  0.8& -212\\
                      &&  \ce{N^{15}NH+}&   3.0(-12)&  292&  0.7&  -69\\
    \ce{N+}&   7.6(-11)&    \ce{^{15}N+}&   1.5(-13)&  504&  1.1& -460\\
  \ce{C2N+}&   8.8(-13)&  \ce{C2^{15}N+}&   3.3(-15)&  271&  0.6&    4\\
 \ce{H2CN+}&   9.6(-12)& \ce{H2C^{15}N+}&   3.8(-14)&  255&  0.6&   67\\
   \ce{NH+}&   4.4(-16)&   \ce{^{15}NH+}&   8.9(-19)&  497&  1.1& -453\\
  \ce{NH2+}&   2.2(-15)&  \ce{^{15}NH2+}&   4.8(-18)&  470&  1.1& -421\\
  \ce{NH3+}&   4.5(-13)&  \ce{^{15}NH3+}&   1.0(-15)&  439&  1.0& -381\\
  \ce{NH4+}&   6.9(-12)&  \ce{^{15}NH4+}&   1.6(-14)&  426&  1.0& -362\\
      \bottomrule
    \end{tabular}
  \end{center}
  \begin{list}{}{}
  \item $\S$: Fractional abundance with respect to \nh. Numbers in
    brackets are powers of ten.
  \item $\dag$: Abundance ratio of the two isotopologues:
    $n(\ce{XN})/n(\ce{X^{15}N})$. For species containing two
    nitrogen atoms, the ratio has been multiplied by two to emphasize
    that a ratio of 220 indeed means no fractionation for such species.
  \item $\ddag$: Fractionation level, or normalized isotopic
    ratio, is the isotopic ratio normalized to the protosolar value
    \nratio=440 from \cite{marty2011}.
  \item $\sharp$ Depletion factor $\delta_{15}=1000\times[{\cal R}_{\rm ATM}/{\cal
      R}-1]$, where the reference value ${\cal R}_{\rm ATM}=272$ is
    the isotopic ratio of the Earth atmosphere.
  \end{list}
\end{table}

\section{Conclusions and perspectives}

Combining observations of \cfifn\ and \thcn\ and assuming a CN:\thcn\ isotopic ratio, we have estimated the CN:\cfifn\ ratio towards two dark clouds. In both cases, the CN molecule appears \phb{slightly} depleted in \fifn\ with respect to the solar wind elemental ratio, with a CN:\cfifn\ \phb{ratio $\approx 500\pm75$ or depletion values $\delta_{15}=[272/\ratio{CN}-1] = -527$ to $-360$\perth}. Towards L1544, additional constraints are provided by the HCN:\hcfifn\ ratio. We performed chemical model calculations, which include ion-neutral fractionation reactions for nitrogen-bearing species. Our gas-phase model predicts significant fractionation of many species. This is in sharp contrast to the former predictions of \cite{terzieva2000}. Even though the model of \cite{charnley2002} also produces fractionation of nitrogen-bearing species, the comparison with our calculations is not direct because the freeze-out of CO onto dust particles is a prerequisite to fractionation in their calculations. In contrast, we note that our model calculations predict fractionation without the need for depletion. Quite interestingly, our model is able to reproduce simultaneously \fifn-poor CN and \fifn-rich HCN, \phb{as is observed towards L1544}. This indeed also demonstrates that these two observational facts do not contradict the chemical argument of HB13, according to which there are two families of nitrogenated species, namely nitriles, and hydrides, which present different sensitivity to fractionation. However, the present work allowed us to explore the mechanisms responsible for the differential fractionation among nitrogenated species and, in particular, to understand how CN can be \phb{slightly depleted} in \fifn\ and at the same time HCN be enriched in \fifn. \phb{The differential fractionation of CN and HCN is at odds with observations in comets \citep{bockelee2008, manfroid2009}, where similar isotopic ratios are obtained for both species. These two observational facts could be reconciled by assuming that CN is a daughter molecule of HCN (\eg\ by photolysis) \citep[for a discussion, see \eg\ ][]{mumma2011}.} Another limitation in our model is brought to light by the isotopic ratios of \ce{N2H+}/\ce{N^{15}NH+} and \ce{N2H+}/\ce{^{15}NNH+} which are predicted below 380, whilst observationally the former is 461$\pm71$. We suggest that observations of \ce{^{15}NNH+} would bring strong constraints on the fractionation processes at work. In particular, whether these two ratios are equal or different would suggest that some of the underlying fractionation reactions present large activation energy. Another possibility, not considered in this work, is that depletion of gas-phase elements, (in particular CO, which is a predator of \ce{N2H+}) plays a significant role for the fractionation of \ce{N2H+}. On more general grounds, the present calculations suggest that a thorough comparison of the Ohio State University (OSU) chemical network with the one used here will potentially provide insights to the chemical processes at work.

From an astrochemical point of view, observations of as many species as possible is most welcome. However, \cfifn\ detection is time demanding, and further progress may rather benefit from observations of H\cfifn\ and H\fifn C towards a larger sample of sources. In addition, observations of \ce{^{15}NH3} would bring \phb{extremely interesting} constraints, but the discrepancy between the results of \cite{lis2010} and \cite{gerin2009} suggests that the interpretation of such observations may not be straightforward. Observations of \ce{^{15}NO} would also be most valuable. However, this molecule presents both fine and hyperfine splittings such that the intensity may be diluted within too many components, making this species \phb{particularly difficult to detect}. The hypothesis of \cite{hilyblant2013ica} could also be pushed forward by observing isotopologues of more complex nitriles, such as \ce{HC3N} and beyond.

\acknowledgements{We acknowledge the comments of the anonymous referee, which helped to improve the scientific quality of the manuscript. Funding for this project was provided by grants from la \textit{R\'egion Rh\^one-Alpes} and from the \textit{Physico-Chimie du Milieu Interstellaire} program. RLG helds a doctoral fellowship from la \textit{R\'egion Rh\^one-Alpes}.}

\bibliographystyle{aa}
\bibliography{general,chemistry,disks,phb,solarsystem,technic}

\appendix

\section{Chemical model}

We implemented the fractionation reactions of TH00, assuming a
symmetry factor $f(B,m)$ of unity for all reactions, unless \ce{N2}
appears as a reactant or as a product. In these cases, $f(B,m)=0.5$ or
1 respectively. A constant Langevin rate of \dix{-9}\cccs\ was adopted
for these ion-neutral reactions. The zero-point energy differences are
taken from TH00. The reactions and their rate coefficients are listed
in Table~\ref{tab:reactions}.

\begin{table}[t]
  \centering
  \caption{Elemental abundances \citep[taken from\ ][]{flower2003}, except for O.
    Numbers in parentheses are powers of 10.}
  \begin{tabular}{l r}
    \toprule
    Species & $n(X)/\nh$\\
    \midrule
    He     & 0.1     \\
    N      & 6.4(-5) \\
    \fifn  & 1.5(-7) \\
    O      & 2.1(-4) \\
    C$^+$  & 8.3(-5) \\
    S$^+$  & 1.9(-6) \\
    Fe$^+$ & 3.0(-9) \\
    \bottomrule
  \end{tabular}\\
  Note: $\nh = n(\h) + 2n(\hh)$
  \label{tab:abinit}
\end{table}

\begin{table*}
  \begin{center}
    \caption{Fractionation reactions and rate coefficients ($k(T) =
      \alpha (T/300)^\beta \exp(-\Delta E /T)$\cccs) implemented in
      our chemical network (from TH00).}
    \label{tab:reactions}
    \begin{tabular}{l c c c r}
      \toprule
      Reaction                         & $\alpha$ & $\beta$ & $\Delta E$ & $k$(10~K)\\
      & \dix{-9}\cccs & & K & \cccs\\
      \midrule
      \ce{$^{15}$NN  +   N2H+    -> N2    +  N$^{15}$NH+} &  1.0 & 0.0 &  0.0 &   1.0(-09)\\
      \ce{N2   +   N$^{15}$NH+   -> $^{15}$NN   +  N2H+ } &  2.0 & 0.0 & 10.7 &   6.9(-10)\\
      \ce{$^{15}$NN  +   N2H+    -> N2    +  $^{15}$NNH+} &  1.0 & 0.0 &  0.0 &   1.0(-09)\\
      \ce{N2   +   $^{15}$NNH+   -> $^{15}$NN   +  N2H+ } &  2.0 & 0.0 &  2.3 &   1.6(-09)\\
      \ce{$^{15}$N   +   N2H+    -> $^{15}$NNH+ +  N    } &  1.0 & 0.0 &  0.0 &   1.0(-09)\\
      \ce{N    +   $^{15}$NNH+   -> $^{15}$N    +  N2H+ } &  1.0 & 0.0 & 36.1 &   2.7(-11)\\
      \ce{$^{15}$N   +   N2H+    -> N$^{15}$NH+ +  N    } &  1.0 & 0.0 &  0.0 &   1.0(-09)\\
      \ce{N    +   N$^{15}$NH+   -> $^{15}$N    +  N2H+ } &  1.0 & 0.0 & 27.7 &   6.3(-11)\\
      \ce{$^{15}$N+  +   N2      -> N+    +  $^{15}$NN  } &  1.0 & 0.0 &  0.0 &   1.0(-09)\\
      \ce{N+   +   $^{15}$NN     -> $^{15}$N+   +  N2   } &  0.5 & 0.0 & 28.3 &   3.0(-11)\\
      \ce{$^{15}$N+  +   NO      -> N+    +  $^{15}$NO  } &  1.0 & 0.0 &  0.0 &   1.0(-09)\\
      \ce{N+   +   $^{15}$NO     -> $^{15}$N+   +  NO   } &  1.0 & 0.0 & 24.3 &   8.8(-11)\\
      \ce{$^{15}$N   +   H2CN+   -> N     +  H2C$^{15}$N+}&  1.0 & 0.0 &  0.0 &   1.0(-09)\\
      \ce{N    +   H2C$^{15}$N+  -> $^{15}$N    +  H2CN+ }&  1.0 & 0.0 & 35.9 &   2.8(-11)\\
      \ce{$^{15}$N   +   H2NC+   -> N     +  H2$^{15}$NC+}&  1.0 & 0.0 &  0.0 &   1.0(-09)\\
      \ce{N    +   H2$^{15}$NC+  -> $^{15}$N    +  H2NC+ }&  1.0 & 0.0 & 35.9 &   2.8(-11)\\
      \bottomrule
    \end{tabular}
  \end{center}
\end{table*}

\section{Collisional rate coefficients}

The hyperfine rate coefficients for $^{13}$CN+H$_2$ and
C$^{15}$N+H$_2$ were derived from the fine-structure rate coefficients
computed by \cite{kalugina2012} for CN+H$_2$. The latter coefficients were
determined from fully quantum close-coupling (CC) calculations based
on a highly correlated potential energy surface. Rate coefficients
were deduced for temperatures ranging from 5 to 100~K. Full details can
be found in \cite{kalugina2012}.

The $^{13}$C (with nuclear spin $I=1/2$) and $^{15}$N ($I=1/2$)
substitutions in CN (in which $^{14}$N has a nuclear spin $I=1$)
significantly modify the hyperfine structure of the molecule. Yet, the
fine-structure rate coefficients can be considered nearly identical
for the three isotopologues. To derive the $^{13}$CN and
C$^{15}$N hyperfine rate coefficients, we used the
infinite-order-sudden (IOS) approximation applied to the CN
fine-structure rate coefficients of \cite{kalugina2012}. Within this
approximation, where the fine-structure energy spacings are ignored
compared to the collision energy, the rate coefficients between
fine-structure levels ($k_{Nj \to N'j'} (T)$ for a $^{2}\Sigma$ state
molecule) can be obtained directly from the ``fundamental''
fine-structure rate coefficients (those out of the lowest $N=0$ level)
as follows:
\begin{eqnarray}
\label{iosrot}
&  & k^{IOS}_{Nj \to N'j'}(T)  =  (2j'+1)\sum_L \frac{2L+1}{L+1} \left(\begin{array}{ccc}
j' & L & j \\ 
-1/2 & 0 & 1/2
\end{array}\right)^{2} \nonumber \\ 
& & \hspace{1cm} \times \frac{1}{2}[1-\epsilon (-1)^{j+j'+L}]
  k^{IOS}_{0,1/2 \to L,L+1/2}(T),
\end{eqnarray}
where $\epsilon$ is equal to +1 if the parity of initial and final
rotational $Nj$ level is the same or -1 if the parity of initial and
final rotational $Nj$ level differ\footnote{Note that there is a typo
  regarding the sign of $\epsilon$ in Eq.~(7) of \cite{faure2012}}.

For C$^{15}$N, which possesses a single non-zero nuclear spin, the IOS
rate coefficients among hyperfine-structure levels can be obtained
from the $k^{IOS}_{0, 1/2 \to L,L+1/2} (T)$ rate coefficients as
\begin{eqnarray}
\label{REEQ2}
& & k^{IOS}_{NjF \to N'j'F'} (T)  =  (2j+1)(2j'+1) (2F'+1)\sum_{L} \frac{2L+1}{L+1}  \nonumber \\
& & \times \left( 
\begin{array}{ccc}
j' & L & j \\
-1/2 & 0 & 1/2 
\end{array}
\right)^2
\left\{
\begin{array}{ccc}
j & j' & L \\
F' & F & I 
\end{array}
\right\}^2 \nonumber \\
& & \times \frac{1}{2}[1+\epsilon (-1)^{j+j'+L}] k^{IOS}_{0,1/2 \to L,L+1/2}(T),
\end{eqnarray}
where $I=1/2$ is the nuclear-spin of $^{15}$N. In practice, the CN
fine-structure energy spacings are not negligibly small and the IOS
approximation is expected to fail at low temperature
($T<100$~K). However, since it correctly predicts the relative rates
among hyperfine levels (because the propensity rules are properly
included through the Wigner coefficients), a simple method to correct
the low temperature results is to scale the IOS results, as originally
suggested by \cite{neufeld1994}:
\begin{equation}
  \label{scaling2}
  k^{INF}_{NjF \to N'j'F'} (T) = \frac{k^{IOS}_{NjF \to N'j'F'}(T)}{k^{IOS}_{Nj\to
      N'j'}(T)}k^{CC}_{Nj\to N'j'}(T).
\end{equation}
In this approach the CC rate coefficients $k^{CC}(0, 1/2 \to L,
L+1/2)$ must be employed as the IOS \emph{fundamental} rates in both
Eq.~\ref{iosrot} and \ref{REEQ2}. The scaling procedure thus ensures
that
\begin{equation}
  \label{hypsum}
  \sum_{F'} k^{INF}_{NjF \to N'j'F'}(T)=k^{CC}_{Nj\to N'j'}(T).
\end{equation}
It should be noted that the propensity rule $\Delta j=\Delta F$
predicted by the recoupling approach is also properly reproduced by
IOS approximation, as discussed in \cite{faure2012}. We note that for CN
there is also a strong propensity for transitions with even $\Delta N$
\cite{kalugina2012}. In practice, the first 22 hyperfine levels of
C$^{15}$N were considered, corresponding to rate coefficients for all
210 transitions among levels with $N\leq 5$.

For $^{13}$CN, which possesses two non-zero nuclear spins, the IOS rate
coefficients among hyperfine structure levels can be obtained
similarly, including an additional coupling:
\begin{eqnarray}
\label{REEQ3}
& & k^{IOS}_{NjF_1F \to N'j'F_1'F'} (T)  =  (2j+1)(2j'+1)(2F_1+1)(2F_1'+1)(2F'+1) \nonumber \\ 
& & \times \sum_{L} \frac{2L+1}{L+1} \left( 
\begin{array}{ccc}
j' & L & j \\
-1/2 & 0 & 1/2 
\end{array}
\right)^2
\left\{
\begin{array}{ccc}
j & j' & L \\
F_1' & F_1 & I_1 
\end{array}
\right\}^2 \
\left\{
\begin{array}{ccc}
j & j' & L \\
F' & F & I_2 
\end{array}
\right\}^2 \nonumber \\ & & \times \frac{1}{2}[1-\epsilon
  (-1)^{j+j'+L}] k^{IOS}_{0,1/2 \to L,L+1/2}(T),
\end{eqnarray}
where $I_1=1/2$ and $I_2=1$ are the nuclear spins of $^{13}$C and
$^{14}$N respectively. The propensity rule is in this case $\Delta
j=\Delta F_1=\Delta F$ and the scaling formula writes
\begin{equation}
  \label{scaling3}
  k^{INF}_{NjF_1F \to N'j'F_1'F'} (T) = 
  \frac{k^{IOS}_{NjF_1F \to N'j'F_1'F'}(T)}{k^{IOS}_{Nj\to
      N'j'}(T)}k^{CC}_{Nj\to N'j'}(T),
\end{equation}
In practice, the first 62 hyperfine levels of $^{13}$CN were
considered, corresponding to rate coefficients for all 1676
transitions among levels with $N\leq 5$.

\begin{figure*}
  \centering
  \includegraphics[width=0.45\hsize]{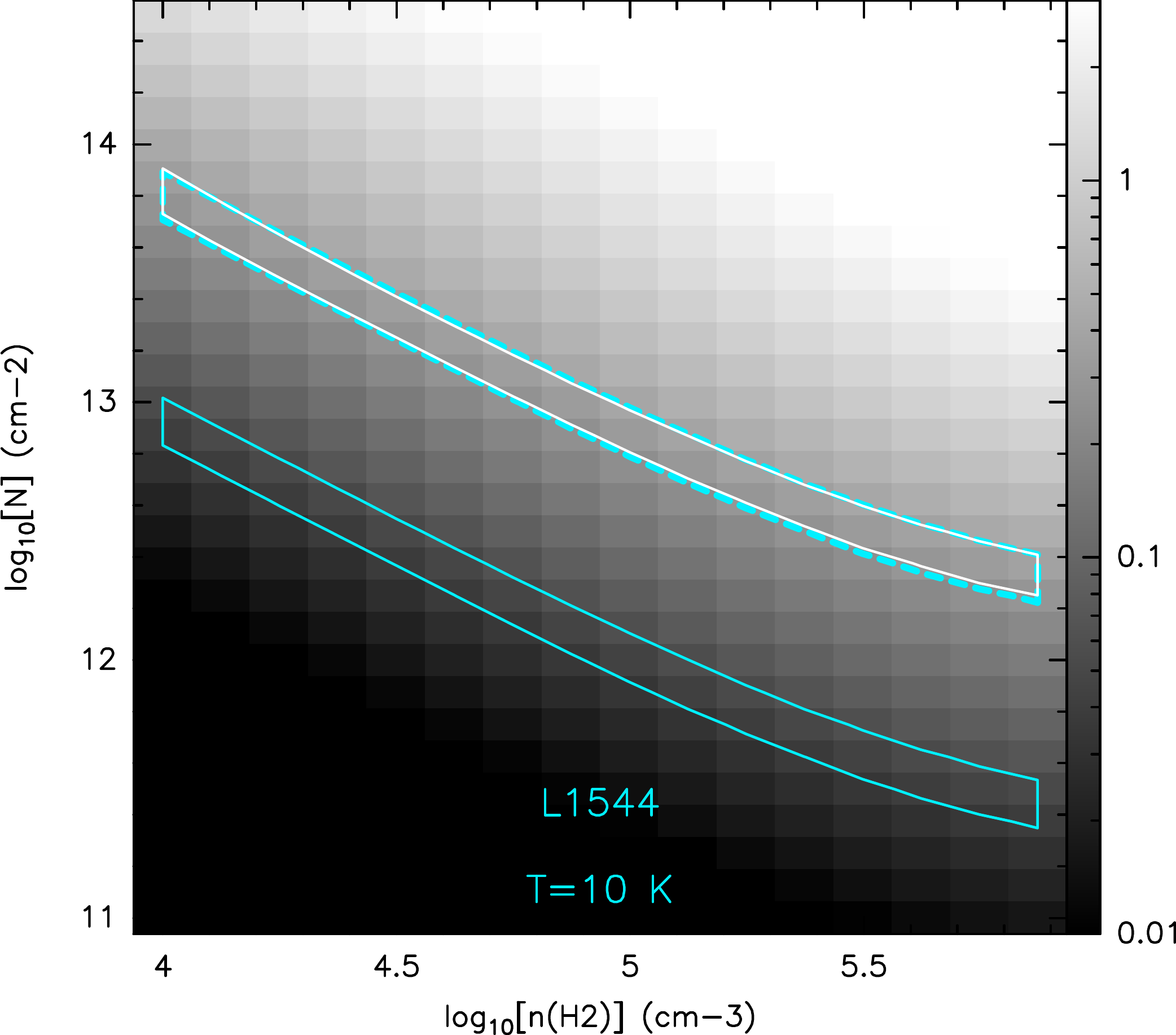}\hfill%
  \includegraphics[width=0.45\hsize]{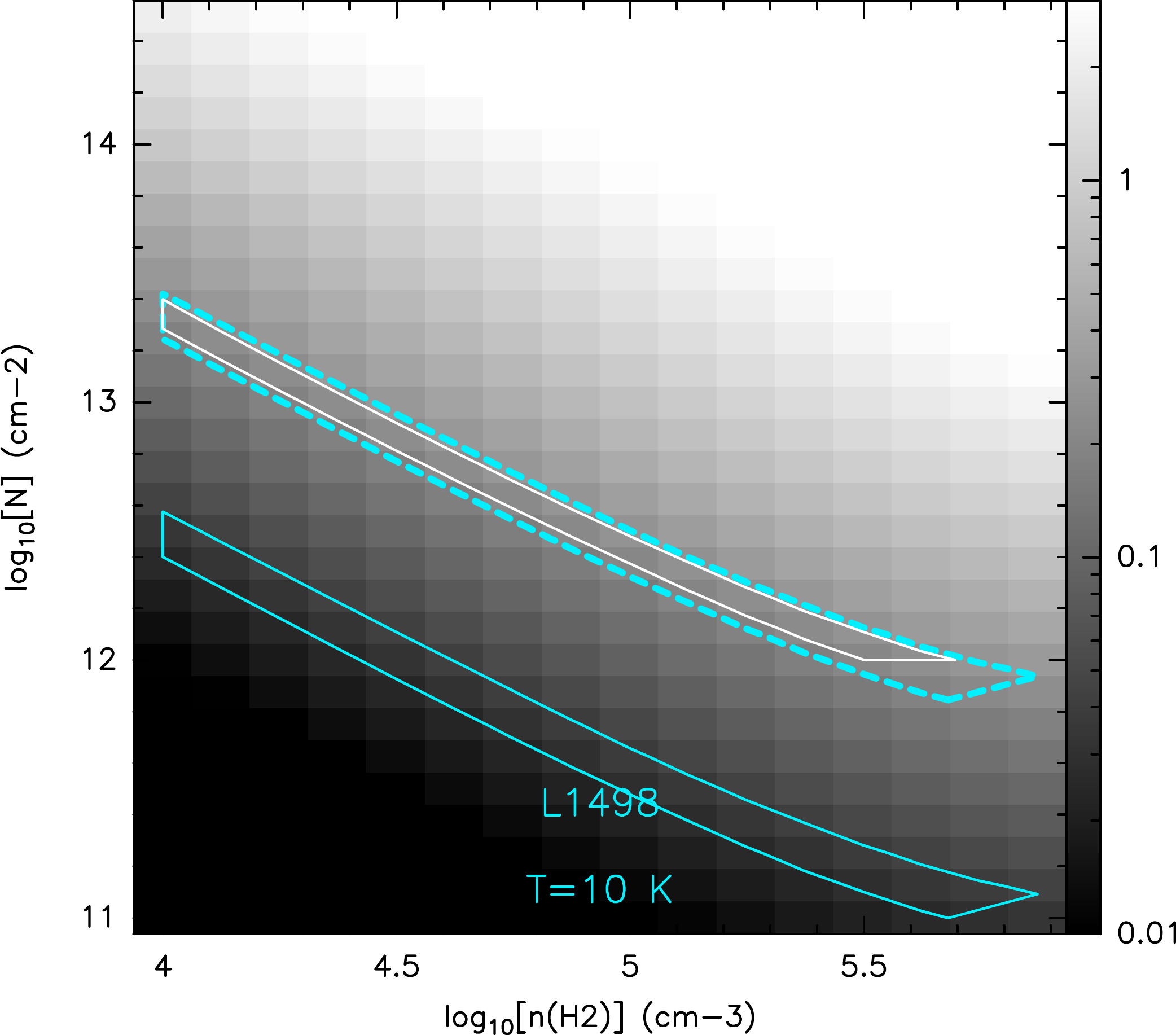}
  \caption{LVG predictions at $T=10$~K towards L1544 (left) and L1498
    (right). In each panel, the grey scale shows the predicted
    intensity of the 110024.590~MHz component of the \cfifn(1-0)
    hyperfine multiplet. The boxes (full line) delineate the solutions
    for the \cfifn\ (cyan) and the \thcn\ lines
    (white). The dashed box shows the \cfifn\ solutions when the \cfifn\ column
    density is multiplied by factors of 7.5 and 7 for L1544 and L1498
    respectively.}
  \label{fig:lvg}
\end{figure*}

\end{document}